\documentclass{article}
\usepackage{arxiv}

\usepackage{amsmath,amssymb}
\usepackage{rotating}
\DeclareMathOperator*{\argmin}{arg\,min}
\usepackage{changepage}
\usepackage[utf8x]{inputenc}
\usepackage{textcomp,marvosym}
\usepackage{cite}
\usepackage{nameref,hyperref}
\usepackage[right]{lineno}
\usepackage{microtype}
\DisableLigatures[f]{encoding = *, family = * }
\usepackage[table]{xcolor}
\usepackage{booktabs}
\usepackage{array}
\newcolumntype{+}{!{\vrule width 2pt}}
\newlength\savedwidth

\usepackage{amsmath, amsfonts, amsthm,amssymb}

\newcommand\thickhline{\noalign{\global\savedwidth\arrayrulewidth\global\arrayrulewidth 2pt}%
\hline
\noalign{\global\arrayrulewidth\savedwidth}}

\usepackage[aboveskip=1pt,labelfont=bf,labelsep=period,justification=raggedright,singlelinecheck=off]{caption}

\usepackage{lastpage,fancyhdr,graphicx}
\usepackage{grffile}
\usepackage{epstopdf}
\fancyheadoffset[L]{2.25in}
\fancyfootoffset[L]{2.25in}
\begin{document}
\vspace*{0.2in}

\begin{flushleft}
{\Large
\textbf\newline{Overlapping  community detection in networks based on link partitioning and partitioning around medoids} % Please use "sentence case" for title and headings (capitalize only the first word in a title (or heading), the first word in a subtitle (or subheading), and any proper nouns).
}
\newline
\\
Alexander Ponomarenko\textsuperscript{1,*},
Leonidas Pitsoulis\textsuperscript{2},
Marat Shamshetdinov\textsuperscript{3}
\\
\bigskip
\textbf{1} National Research University Higher School of Economics, Nizhny Novgorod, Russia
\\
\textbf{2} University of Thessaloniki, Greece
\\
\textbf{3} Intel Corp, Nizhny Novgorod, Russia
\\
\bigskip

% Insert additional author notes using the symbols described below. Insert symbol callouts after author names as necessary.

%\textcurrency Current Address: Dept/Program/Center, Institution Name, City, State, Country % change symbol to "\textcurrency a" if more than one current address note
% \textcurrency b Insert second current address
% \textcurrency c Insert third current address

% Use the asterisk to denote corresponding authorship and provide email address in note below.
* aponomarenko@hse.ru

\end{flushleft}
\section*{Abstract}
In this paper, we present a new method for detecting overlapping communities in networks with a predefined number of clusters called LPAM (Link Partitioning Around Medoids).
The overlapping communities in the graph are obtained by detecting the disjoint communities in the associated line graph employing link partitioning and partitioning around medoids which are done through the use of a distance function defined on the set of nodes.
We consider both the commute distance and amplified commute distance as distance functions.
The performance of the LPAM method is evaluated with computational experiments on real life instances, as well as synthetic network benchmarks.
For small and medium-size networks, the exact solution was found, while for large networks we found solutions with a heuristic version of the LPAM method.

%%%%%%%%%%%%%%%%%%%%%%%%%%%%%%%%%%%%%%%%%%%%%%%%%%%%%%%%%%%%%%%%%%%%%%%%%%%%%%%%%%%%%%%%%%%%%%%%%%%%%%%%%%%%%%%%%%%%%%%%%%%%%%%%%%%%%%%%%%%%%
%%   Introduction
%%%%%%%%%%%%%%%%%%%%%%%%%%%%%%%%%%%%%%%%%%%%%%%%%%%%%%%%%%%%%%%%%%%%%%%%%%%%%%%%%%%%%%%%%%%%%%%%%%%%%%%%%%%%%%%%%%%%%%%%%%%%%%%%%%%%%%%%%%%%%
\section*{Introduction}
Detection of overlapping communities in a network is the task of grouping the nodes of the network into a family of subsets called clusters so that each cluster contains nodes that are similar with respect to the overall network structure. Overlapping means that one node can belong to multiple clusters, in contrast to disjoint community detection where the clusters form a partition of the set of nodes.
To this day, there is no widely accepted formal definition for the notion of community in a network.
This leads to different community definitions and allows for the existence of a variety of graph clustering methods that can only be compared based on their computational complexity and the empirical evaluation of their proposed communities.
A common approach to formalizing the notion of community in a network is through the use of quality functions that attempt to quantify the degree of the community structure captured by a given partition of the nodes.
That is, a quality function will, in principle, attain extreme values for the clustering of the nodes that best reflects the community structure of a graph.
Given such a quality function, community detection translates into an optimization problem.
Modularity~\cite{Girvan:2004} is one of the best-known quality functions.
It has been used by many methods that solve the related optimization problem with varying success.
However, it is still an open question what the properties of a good quality function are~\cite{laarhoven:2014}.

Community detection in networks is an actively developing area connected to many fields of science that need tools for complex network analysis, including molecular biology, sociology, data mining, and unsupervised machine learning.
Network clustering methods can be classified according to the approaches on which they are based.
%%%%%%%%%%%%%%%%%%%%%%%%%%%%%%%%%%%%%%%%%%%%%%%%%%%%%%%%%%%%%%%%%%%%%%%%%%%%%%%%%%%%%%%%%%%%%%%%%%%%%%%%%%%%%%%%%%%%%%%%%%%%%%%%%%%%%%%%%%%%%

There is a plethora of different methods and approaches for overlapping community detection in graphs, which can be partially attributed to the absence of a well defined and widely accepted quality function for overlapping communities, as is the case with non-overlapping community detection. An attempt to axiomatize quality functions for non-overlapping graph clustering, in the form of
intuitive properties that any such function should satisfy, is presented in~\cite{laarhoven:2014}. The authors,
driven by similar results in distance-based clustering, propose six such properties. For instance, the value of a clustering quality function should not decrease if, for a given clustering, we add edges between nodes in the same clusters.
Moreover, they showed that modularity does not satisfy some of these properties.

In a more recent work~\cite{Chakraborty_etal:2017}, the authors compiled a survey of the currently known families of quality functions for both non-overlapping and overlapping graph clustering.
The authors also present computational experiments on a set of benchmark instances with known community structure, then compare the quality functions in terms of performance in identifying the communities.
The most recent overview and classification of the state-of-the-art methods for overlapping community detection, as well as a computational comparison of existing solutions and benchmark instance evaluation, can be found in~\cite{xie:2013overlapping}.
In the paper, the authors present 14 different algorithms and propose a unified framework for testing them.

One approach to overlapping community detection is link partitioning, also known as link communities identification, which involves splitting a set of edges instead of partitioning nodes.
It is based on the idea that relations between nodes define the community structure, not the nodes themselves.
In the case of link partitioning, a node belongs to a community if it has adjacent edges that belong to that community.
For example, a person may play soccer with a group of playmates at weekends and go to work with coworkers on weekdays. Given that a coworker can also be a playmate, we have overlapping communities.
That person has two types of relations with other people: "plays soccer with" and "works with." Thus, the person belongs to two communities: "soccer players" and "colleagues."
Thereby, we can consider this person as an overlapped node.

Even though link partitioning for overlapping community detection seems very natural, historically, the methods that exploit it appeared relatively late.
In 2009 and later, in 2010, Evans and Lambiotte \cite{PhysRevE.80.016105}, \cite{evans2010line} were the first who performed node partitioning of a line graph to obtain an edge partition of the original graph.
So they projected the network into a weighted line graph whose nodes are the links of the original graph, and, after that, they applied one of the disjoint community detection algorithms.
In 2011, Kim and Jeong \cite{kim2011map} proposed a modified version of the map equation method (also known as Infomap \cite{rosvall2008maps} ) to detect link communities under the Minimum Description Length (MDL) principle.
Also, Evans \cite{evans2010clique} in 2010, extended the line graph approach to using clique graphs, wherein cliques of a given order are represented as nodes in a weighted graph.
%%spectral clustering approach

To reveal overlapping communities, calculating the spectrum of a graph with a Laplacian matrix was also exploited in some papers.
Selecting the eigenvectors corresponding to the first k smallest eigenvalues of the graph, the Laplacian matrix allows embedding each node into a k-dimensional space, with the expectation that nodes from the same cluster will have a small distance to one another, relative to nodes outside the cluster.
To apply the spectral approach to revealing the overlapping community structure, the authors in~\cite{zhang2014detecting} suggest using the K-medians algorithm, instead of the regular K-means, for clustering in the k-dimensional spectral domain.
A Gaussian Mixture Model in the spectral domain was proposed in~\cite{MagdonIsmail2011SSDEClusterFO}, and a fuzzy $c$-means algorithm, to obtain a soft assignment, was used in~\cite{zhang2007identification}.

%Indeed the link partitioning approach also can be applied in a combination with a spectral approach to finding overlapping communities however currently there is no study about such combination by our knowledge.

%Spectral clustering mainly based on the observation that Laplacian matrix of a graph with several

%%%%%%%%%%%%%%%%%%%%%%%%%%%%%%%%%%%%%%%%%%%%%%%%%%%%%%%%%%%%%%%%%%%%%%%%%%%%%%%%%%%%%%%%%%%%%%%%%%%%%%%%%%%%%%%%%%%%%%%%%%%%%%%%%%%%%%%%%%%%%
%%   Organization of paper

In this paper, we present a novel approach to overlapping community detection.
The paper is organized as follows:
In \nameref{sec:material_methods}, we give a formal definition of the clustering problem, describe the proposed method and the datasets, and compare the methods used in computational experiments.
We briefly discuss how to choose the input parameters and give the estimation of the computational cost of the proposed method in section \nameref{sec:Discussion}.
Traditionally, we finish the paper with the \nameref{sec:Conclusion} section.

%%%%%%%%%%%%%%%%%%%%%%%%%%%%%%%%%%%%%%%%%%%%%%%%%%%%%%%%%%%%%%%%%%%%%%%%%%%%%%%%%%%%%%%%%%%%%%%%%%%%%%%%%%%%%%%%%%%%%%%%%%%%%%%%%%%%%%%%%%%%%
%%  Materials and Methods
%%%%%%%%%%%%%%%%%%%%%%%%%%%%%%%%%%%%%%%%%%%%%%%%%%%%%%%%%%%%%%%%%%%%%%%%%%%%%%%%%%%%%%%%%%%%%%%%%%%%%%%%%%%%%%%%%%%%%%%%%%%%%%%%%%%%%%%%%%%%%

\section*{Materials and Methods}\label{sec:material_methods}

%%%%%%%%%%%%%%%%%%%%%%%%%%%%%%%%%%%%%%%%%%%%%%%%%%%%%%%%%%%%%%%%%%%%%%%%%%%%%%%%%%%%%%%%%%%%%%%%%%%%%%%%%%%%%%%%%%%%%%%%%%%%%%%%%%%%%%%%%%%%%
\subsection*{Problem statement}
Let $G=(V,E)$ be a graph with $n$ nodes $V = \{v_1, v_2,...,v_n\}$ and $m$ edges $E \subseteq V \times V $. For a given natural number $k$
define a
\textit{cover} as a family of $k$ subsets of nodes
\[
  \mathcal{C} = \{ C_1, C_2, \ldots, C_k \}
  \]
  where each $C_i$ is called a \textit{cluster} or \textit{community}. The goal in community detection is to find a cover $\mathcal{C}$ which
best describes the community structure of the graph, in the sense that nodes within clusters are more densely connected than the clusters
themselves.
We can also associate with $\mathcal{C}$ an \textit{affiliation matrix} $\mathbf{F}_{\mathcal{C}} \in \mathbb{R}^{|V|\times |\mathcal{C}|}$ where $F_{vc}$ corresponds to the degree of affiliation of vertex $v$ with community $c \in \mathcal{C}$. If we impose the following constraints
\begin{eqnarray}
  \sum_{c\in\mathcal{C}} F_{vc} = 1, & & \forall v \in V \\ \label{eq:1}
  0 \leq F_{vc} \leq 1, & & \forall v \in V, \forall c \in\mathcal{C}. \label{eq:2}
\end{eqnarray}
then the values of the affiliation matrix are also known as \textit{belonging coefficients}~\cite{Shen_2009}.
In the case of non-overlapping community detection we have that $\mathcal{C}$ must be a partition of $V$, or equivalently, equation \eqref{eq:2} is\
replaced by the binary constraint $F_{vc} \in \{0,1\}$.
%%%%%%%%%%%%%%%%%%%%%%%%%%%%%%%%%%%%%%%%%%%%%%%%%%%%%%%%%%%%%%%%%%%%%%%%%%%%%%%%%%%%%%%%%%%%%%%%%%%%%%%%%%%%%%%%%%%%%%%%%%%%%%%%%%%%%%%%%%%%%\subsection*{Proposed method}\label{sec:proposed_method}
The proposed method is based on non-overlapping link partitioning. Thus, the task of overlapping community detection is reduced to the problem of finding non-overlapping communities in the set of edges, which is equivalent to the problem of finding non-overlapping communities on a line graph $L(G)$ whose vertices correspond to edges of the original graph $G$. Two vertices are connected by an edge in $L(G)$ if the corresponding edges in $G$ have a common node.
In order to determine disjoint communities in the line graph $L(G)$ we build a distance matrix $D = (d_{ij}) \in \mathbb{R}^{m\times m}$ based on the structure of $L(G)$, and for doing this we utilize a distance function on the nodes of a graph.
For this purpose we tested two distance functions; the commute distance~\cite{yen2005commuteDistance} and the amplified commute distance~\cite{luxburg2010}.
Given that we seek to find $k$ overlapping communities in the original graph $G$, we compute a set $S = \{s_1, s_2, ..., s_k\}$ of vertices from $L(G)$ which can be considered the medians with respect to the distances in $D$, that is
\begin{equation}\label{eq:p_median_1}
  S := \argmin_{T \subset E(G), |T|=k} \sum_{j\in E(G)} \sum_{c\in T} {d_{jc}x_{jc}},
\end{equation}
\begin{align}\label{eq:p_median_2}
  &    \begin{matrix}
     x_{jc} & =
      & \left\{
      \begin{matrix}
      1, & \mbox{if}\;\;\; d_{jc} \leq  d_{js}, \: s \in S,  \\
      0, & \mbox{otherwise }
      \end{matrix} \right.
      \end{matrix}
  \end{align}
Thus, $x_{jc}$ is an indicator variable which takes the value 1 when the edge $j$ of the original graph $G$ belongs to cluster $c$ and 0 otherwise.
The $\argmin$ function runs over all possible subsets $T$ of $E(G)$ of the size $k$.
Together, expressions \ref{eq:p_median_1} and \ref{eq:p_median_2} constitute the $k$-median problem also known as the \textit{facility location problem} which is known to be NP-complete \cite{fowler1981optimal, gonzalez1985clustering}.

The matrix of belonging coefficients for the final covering is calculated as follows.
\begin{align}
  \label{eq:final_covering}
  &    \begin{matrix}
     F_{ic} & =
      & \left\{
      \begin{matrix}
      1, & \mbox{if} \;\;\;\frac{\sum_{(i,j) \in E} {x_{jc}}      }{d_i} \geq \theta,  \\
      0, & \mbox{otherwise }
      \end{matrix} \right.
      \end{matrix}
  \end{align}
for every $i\in V(G)$ and $c\in S$, where $\theta$ is a threshold parameter, and $d_i$ is the degree of vertex $i$ in the graph $G$.
Thus, the belonging coefficient $F_{ic}$ of node $i$ to cluster $c$ is proportional to the number of adjacent edges belonging to the cluster $c$. As the value of $\theta$ increases, the degree of overlapping between the communities also increases.

In summary, in order to find $k$ overlapping communities in a graph $G$, our proposed method {\bf L}ink {\bf P}artitioning {\bf A}round {\bf M}edoids (LPAM)consists of the following steps:
\begin{enumerate}
  \item Build the line graph $L(G)$
  \item Find $k$ disjoint communities in $L(G)$:\begin{enumerate}
    \item Compute the distance matrix $D$ between each pair of nodes based on commute distance or amplified commute distance
    \item Solve the $k$-median problem based on the distances in $D$ and compute the medians $S = \{s_1, s_2, ..., s_k\}$
  \end{enumerate}
  \item Build a cover for the original graph $G$ based on the affiliation matrix $\mathbf{F}_{\mathcal{C}}$ which is constructed from $S$ and a threshold value $\theta$.
\end{enumerate}

The value of the parameter $\theta$ should be chosen according to the application.
In our experiments, we have found that in most cases a threshold value of $0.5$ for $\theta$ produces the best result.
The main reason for this is that most networks with known ground truth cluster assignment have just a few nodes belonging to more than two communities.
When $\theta$ is large, the algorithm tends to assign nodes to a single cluster.
Conversely, with small values of $\theta$, nodes are assigned to many clusters.

It must be noted that the LPAM method is applicable only to unweighted graphs.
Since the method works with a line graph where the nodes are the edges of the original graph, it does not take into account the edge weights of the original graph.

%%%%%%%%%%%%%%%%%%%%%%%%%%%%%%%%%%%%%%%%%%%%%%%%%%%%%%%%%%%%%%%%%%%%%%%%%%%%%%%%%%%%%%%%%%%%%%%%%%%%%%%%%%%%%%%%%%%%%%%%%%%%%%%%%%%%%%%%%%%%%
\subsection*{Distance functions}

There are various options when it comes to choosing a distance function on the nodes of a graph.
Intuitively, a proper distance function should reflect the relationship between nodes within the same cluster in a community, so that the vertices from the same cluster should have a shorter distance between one another than the distance between them if they were to belong to different clusters.
In this paper, we employed two distance functions, namely commute distance and amplified commute distance.

\subsubsection*{Commute distance}
Commute distance~\cite{Lovasz1996} is also known as resistance distance~\cite{klein1993resistance} in the literature.
The resistance distance can be thought of as the effective resistance between two nodes in a graph if we consider this graph to be an electrical circuit. It is defined as
\begin{equation}
  \label{eq:resistance_distance}
  d_\text{r}(i,j)=\frac{K^{(i,j)}}{K_{(ij)}},
\end{equation}
where $K_{(ij)}$ is the minor of a Kirchhoff matrix, and $K^{(i,j)}$ is a second-order algebraic complement, that is, a determinant of the matrix obtained from the Kirchhoff matrix by deleting two rows and two columns $i, j$.
Commute distance $d_{\text{cm}}(i,j)$ and resistance distance $d_{\text{r}}(i,j)$ are related as follows
\begin{equation}
  \label{eq:resistance_and_commute}
  d_\text{cm}(i,j)=\text{vol(G)} d_{\text{r} }(i,j),
\end{equation}
where $\text{vol}(G)=\sum_{v\in V(G)} d_{v}$ is the volume of the graph $G$, and $d_{v}$ is the degree of vertex $v$.
The value of commute distance $d_{\text{cm} }(v,w)$ between node $v$ and node $w$, on the graph $G$, can be interpreted as the expected number of steps that a random walker needs to take to reach node $w$ from $v$ and return back.
Intuitively, commute distance seems like a good candidate for capturing the community structure in a graph, in the sense that nodes within the same community should have a higher probability to be reachable between each other than nodes from different communities.
The number of possible paths between two nodes is directly proportional to the commute distance between these two nodes, and one should expect that pairs of nodes, within the same community, should have a higher number of paths than pairs from different communities.
However, commute distance is theoretically flawed when it comes to large graphs.
In ~\cite{luxburg2010}, the authors proved that when the size of the graph becomes sufficiently large, the probability of reaching a node from another node becomes dependent only on the degree of the destination node.
The authors called this effect {\em losing in space}.
To overcome this drawback, the authors proposed amplified commute distance as a possible improvement in the same paper.
\subsubsection*{Amplified commute distance}Amplified commute distance can be expressed as
\begin{equation}
   \label{eq:amplified_commute_distance}
   d_\text{amp}(i,j)=\frac{d_{\text{cm} }(i,j)}{\text{vol(G)}} - \frac{1}{d_i} - \frac{1}{d_j} + \frac{2w_{ij}}{d_i d_j } - \frac{w_{ii}}{d^2_i} - \frac{w_{jj}}{d^2_j},
  \end{equation}
  where the purpose of the negative terms is to reduce the influence of the edges adjacent to $i$ and $j$, which completely dominate the behavior of resistance distance. The term {\em amplify} is intended to emphasize the main role of the first term. As well as the original commute distance, the amplified commute distance is Euclidean~\cite{luxburg2010}.

%%%%%%%%%%%%%%%%%%%%%%%%%%%%%%%%%%%%%%%%%%%%%%%%%%%%%%%%%%%%%%%%%%%%%%%%%%%%%%%%%%%%%%%%%%%%%%%%%%%%%%%%%%%%%%%%%%%%%%%%%%%%%%%%%%%%%%%%%%%%%
\subsection*{Benchmarks and evaluation}
To evaluate the quality of the tested algorithms, we employ an implementation of the Normalized Mutual Information measure for sets of overlapping clusters (ONMI)~\cite{McDaidNMI}.
We used it to measure the difference between the covering produced by the examined algorithm and the known ground truth.
In recent literature, ONMI values have become one of the most widely used measures to calculate the difference between two coverings. Since many papers in overlapping clustering (e.g. \cite{xie:2013overlapping, gates2017comparing, cheraghchi2017mining})include ONMI values for comparison purposes with benchmark graph instances with known ground truth, we can evaluate our proposed method without necessarily implementing other methods as long as we use the same benchmark instance set.

\subsubsection*{lattice 8x8 example }
\begin{figure}[h]
  \centering
  \includegraphics[scale=0.25]{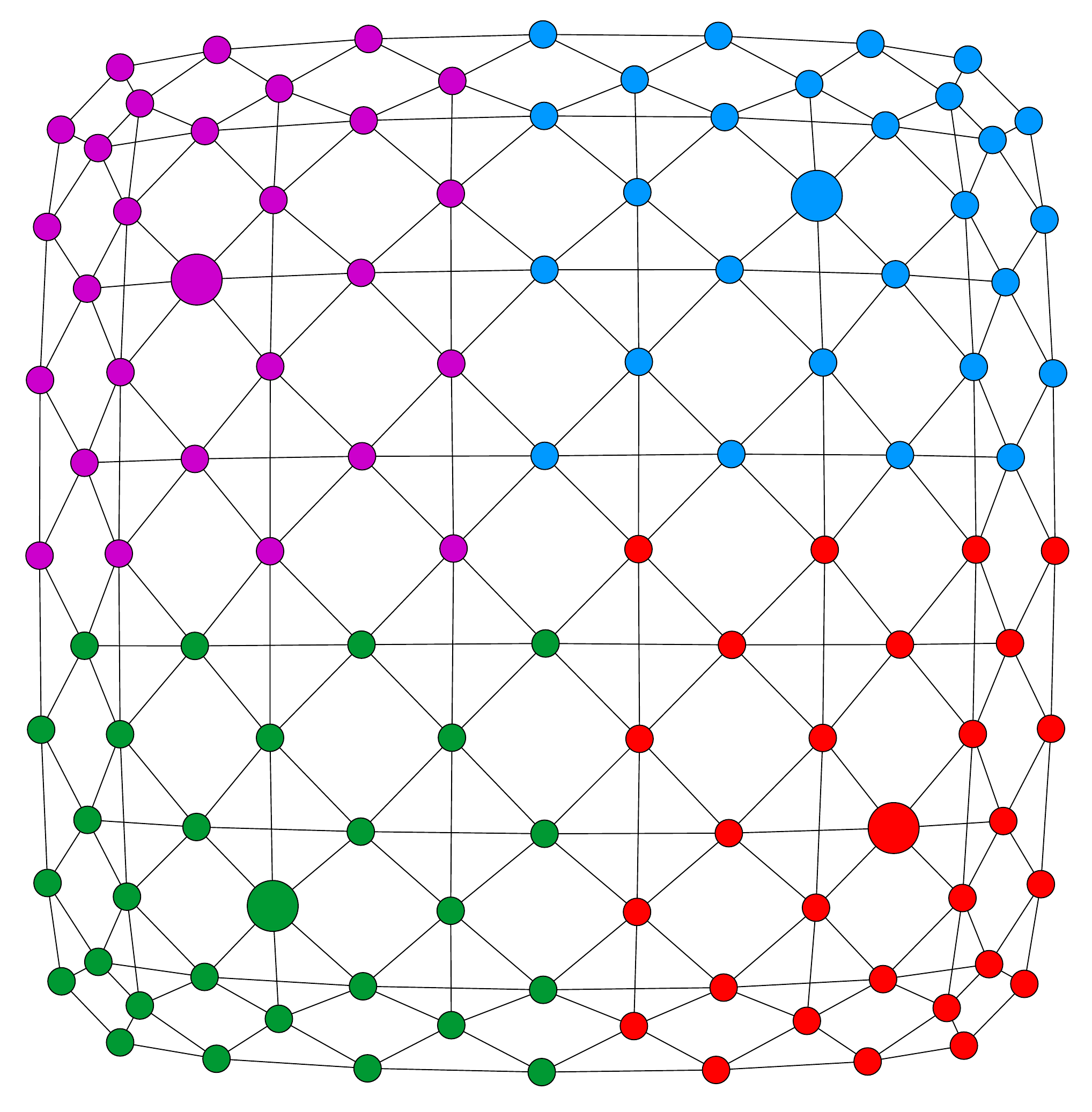}
  \includegraphics[scale=0.25]{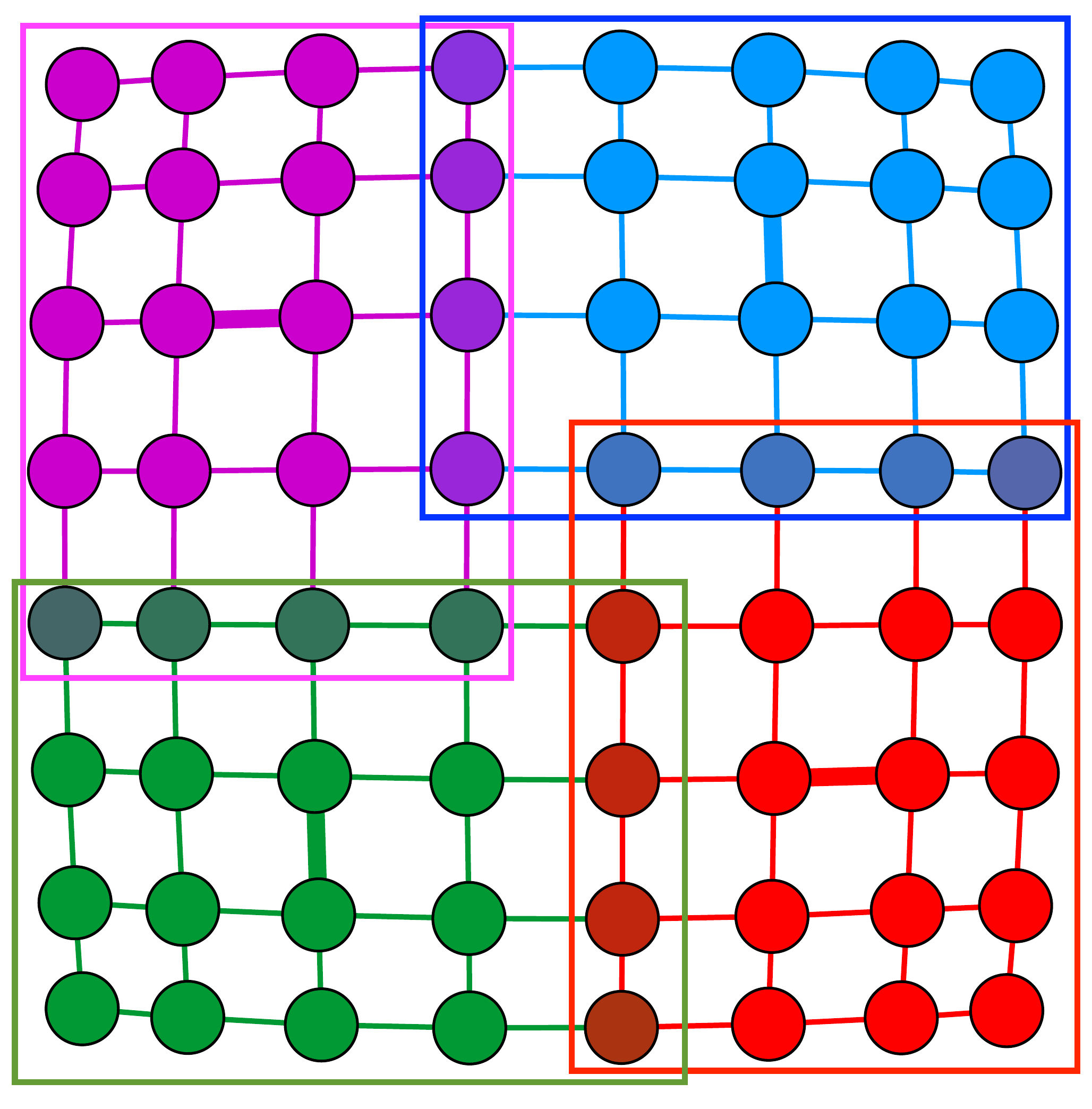}
\caption{The $8\times 8$ lattice on the right and its line graph on the left. 4 clusters are highlighted by the rectangles with corresponding colours. The medoids on the line graph are presented by big circles, while in the original graph medoids are identified by bold edges.} \label{fig:lattice_8x8}
\end{figure}

To help readers gain some intuition into the proposed method, we created a pedagogical example
illustrated in Figure~\ref{fig:lattice_8x8}.
Given a regular $8\times 8$ lattice, which naturally does not contain any community structure, we applied our method for $k=4$ overlapping communities.
As can be seen in Figure~\ref{fig:lattice_8x8}, our method, which produced the same results for both commute and amplified commute distance, identifies four equal and overlapping communities in such a way that each community overlaps exactly with two others.
The medoids on the line graph are presented by big circles, while the communities in the lattice are identified with colors and the corresponding medoids with bold edges. Similarly, if we choose $k=2$, we get two equal overlapping communities.

\subsubsection*{Compared Methods}
Although the performance of the proposed method could in principle be compared with other published methods based solely on the ONMI value, for the sake of consistency we used the publicly available implementations of the following three overlapping community detection methods.

\textbf{Greedy Clique Extension} \cite{lee2010detecting}.
The Greedy Clique Extension algorithm (GCE) can be considered a heuristic for the optimization problem of finding community structures according to the Lancichinetti community quality function ~\cite{lancichinetti2009detecting} $F_S$.
\begin{equation} \label{eq:lancichinetti2009}
  F_S=\frac{k^S_{in}}{(k^S_{in} + k^S_{out})^\alpha},
\end{equation}where $k^S_{in}$ is internal degree and external degree is $k^S_{out} $ of graph subset $S$.
The algorithm has four stages.
\begin{enumerate}
  \item The set of maximum cliques is obtained by using some heuristic algorithms. The cliques with the size of at least $r$ are considered seeds;
  \item Each seed is expanded by a greedy algorithm according to the quality function $F_S$. For a certain seed the expanding process continues until quality function $F_S$ is increasing;
  \item The expanded seeds are merged with the help of the symmetric distance function defined between a pair of communities as
  \begin{equation} \label{eq:comsim}
    \delta_E(S, S')=1 - \frac{S \cap S' }{\min(|S|,|S'|)}.
  \end{equation}
  Communities $S$ and $S'$ are merged if distance $ \delta_E(S, S')$ is less than threshold $\epsilon$. The authors recommend using the values $r = 4$ and $\epsilon = 0.25$.
  The overlapping naturally appears because one vertex can be covered by several seed extensions.

\item Calculate the final communities covering.
% TODO a few words about the method complexity and deterministic/non-deterministic nature of the algorithm.
\end{enumerate}

\textbf{Order Statistics Local Optimization Method} (OSLOM) \cite{lancichinetti:2011}.
The main feature of the OSLOM is that it utilizes the statistically significant community measure as the fitness function.
In turn, a statistically significant community is defined with the help of the configuration model~\cite{molloy1995critical} as a null hypothesis.
%This method uses the metric of the importance of the cluster.
Similar to GCE, the OSLOM is based on optimizing the fitness function by local search, via adding or removing vertices from the cluster.
At the final stage, the OSLOM builds a hierarchical clustering structure. Every cluster is considered as a single vertex.
New vertices are connected if the corresponding clusters have common edges.
The weights of the new edges are assigned proportionally to the number of the edges between the original clusters.

Summarizing, the algorithm has the following steps:
\begin{enumerate}
  \item Find clusters via local search to maximize the fitness function. Repeat until convergence;
  \item Unite or split clusters based on their internal structure;
  \item Consider clusters as vertices. Build a hierarchical structure of clusters iteratively.
\end{enumerate}
%This method can define overlapping clusters, as well as build clusters hierarchies.
%\item \textbf{COPRA} \cite{wu2012balanced}. An iterative method, based on the idea of multi-label propagation with computation complexity close to linear.

\textbf{COPRA} \cite{gregory:2009finding}. This is an iterative method, based on the idea of multi-label propagation with computation complexity close to linear.
It extends the label propagation algorithm (LPA) \cite{zhur:2002learning} with the ability for every node to have multiple labels.
One of the several drawbacks of COPRA is that the node can belong at most to the fixed number of communities $v$ which is a parameter of the algorithm.
To avoid this problem, the BMLPA method was proposed \cite{wu2012balanced}, however, the authors do not provide an implementation, which makes it hard to perform comparison with this method.
The non-deterministic nature of the COPRA algorithm with high variance is another drawback which makes it hard to interpret the results.
Mainly, the randomness comes from two factors. The first is assigning the labels randomly at the initial stage.
The second is part of the label propagation process. If multiple labels have the same maximum belonging coefficient below the threshold, COPRA retains one of them, chosen at random.

Here below we will briefly introduce the remaining methods involved in the comparison. For the consistency

\textbf{PercoMVC} The PercoMVC approach consists of the two steps \cite{kasoro2019percomcv}. In the first step, the algorithm attempts to determine all communities
that the clique percolation algorithm may find. In the second step, the algorithm computes the Eigenvector Centrality method on the output of the first step to measure
the influence of network nodes and reduce the rate of the unclassified nodes

\textbf{danmf} The procedure uses telescopic non-negative matrix factorization in order to learn a cluster membership distribution over nodes. The method can be used in an overlapping and non-overlapping way \cite{ye2018deep}.

\textbf{SLPA} is an overlapping community discovery that extends tha LPA \cite{xie2011slpa}. SLPA consists of the following three stages: 1) the initialization 2) the evolution 3) the post-processing

\textbf{egonet splitter} The method first creates the egonets of nodes. A persona-graph is created which is clustered by the Louvain method \cite{epasto2017ego}.

\textbf{Demon} is a node-centric bottom-up overlapping community discovery algorithm.
It leverages ego-network structures and overlapping label propagation to identify micro-scale communities that are subsequently merged in mesoscale ones \cite{coscia2014uncovering}.

\textbf{k-clique} \cite{palla2005uncovering} Find k-clique communities in graph using the percolation method.
A k-clique community is the union of all cliques of size k that can be reached through adjacent (sharing k-1 nodes) k-cliques.

\textbf{LAIS2} is an overlapping community discovery algorithm based on the density function.
The algorithm considers the density of a group as defined as the average density of the communication exchanges between the actors of the group.
LAIS2 IS composed of two procedures LA (Link Aggregate Algorithm) and IS2 (Iterative Scan Algorithm). \cite{baumes2005efficient}

\textbf{Angel} is a node-centric bottom-up community discovery algorithm.
It leverages ego-network structures and overlapping label propagation to identify micro-scale communities that are subsequently merged into mesoscale ones.
Angel is the, faster, successor of Demon \cite{rossetti2019exorcising}.

The \textbf{Leiden} algorithm is an improvement of the Louvain algorithm \cite{traag2019louvain}. The Leiden algorithm consists of three phases:
\begin{enumerate}
    \item local moving of nodes,
    \item refinement of the partition
    \item aggregation of the network based on the refined partition, using the non-refined partition to create an initial partition for the aggregate network.
\end{enumerate}

The \textbf{Label Propagation algorithm} (LPA) detects communities using network structure alone \cite{raghavan2007near}. The algorithm doesn’t require a pre-defined objective function or prior information about the communities.
It works as follows:
\begin{enumerate}
  \item Every node is initialized with a unique label (an identifier)
  \item These labels propagate through the network
  \item At every iteration of propagation, each node updates its label to the one that the maximum numbers of its neighbours belong to. Ties are broken uniformly and randomly.
  \item LPA reaches convergence when each node has the majority label of its neighbours.
\end{enumerate}

\subsubsection*{Datasets}\label{datasets}

As real-world datasets, we used the following four well-known network instances with the known ground truth.
\begin{itemize}
\item School Friendship: a high school friendship network where the ground truth is a total of 6 communities \cite{ding2016overlapping}.
\item Zachary's karate club: a social network of friendships between 34 members of a karate club at a US university in the 1970s \cite{Zachary:1977}.
\item Word adjacencies: an adjacency network of common adjectives and nouns in the novel David Copperfield by Charles Dickens \cite{Newman:2006}
\item Books about US politics: A network of books about US politics published around the time of the 2004 presidential election and sold by the online bookseller Amazon.com. The edges between the books represent frequent co-purchasing of books by the same buyers. The dataset can be found on Valdis Krebs' website \url{http://www.orgnet.com}
\item American College Football: Graph of the games between college football teams which belong to 12 different confederations \cite{Girvan:2002}.
\end{itemize}

We constructed synthetic networks of the several types with known ground truth.
The graphs created by recently published network generator "FARZ" \cite{fagnan2018modular} have prefix "farz".
A float number in the name is a value for parameter $beta$. For all cases we used $n=200$, $m = 5$, $k=5$, $alpha=0.2$, $gamma=0.5$.
Planted partition graphs (PP-graphs) \cite{fortunato2010community} have prefix "PP".
In turn, the networks (SBM-graph) generated by stochastic block model \cite{holland1983stochastic} have prefix "SBM".
The first float number in the name of SBM/PP-graphs is an intra cluster probability for the edge.
The second is a probability for an edge to be between clusters.
We use prefix {\tt bench} for Lancichinetti networks \cite{Lancichinetti:2008}.
A float number after prefix means mixing parameter (mu). The remaining parameters are $N=200$, $k = 15$, $ \text{maxK}=50$, $ \text{minC}=5$, $ \text{maxC} = 50$, $on= 20$, $om=2$.
% For historical reasons, we have
The parameters used for generating networks bench\_30,..., bench\_60\_dense can be found in \nameref{S8_BENCH_FLAGS}.
All instances of generated graphs can be found in our github repo. In addition, all parameters that were used for the graphs generation can be found inside a script {\tt exp\_CDLIB.py}.
Finally, all values of random seeds in experiments were fixed, thus all computational results can be reproduced.

All the relevant information for the above mentions benchmark networks presented in the Table~\ref{tab:datasets}.

\begin{table}[!ht]
\begin{adjustwidth}{0in}{0in}
\caption{Basic statistics of the networks used in the computational experiments }\label{tab:datasets}
%\centering
\resizebox{7cm}{!}{
\begin{tabular}{lrrrrrrrrr}
    \toprule
    statistic &  $\hat{c}$ &  $\hat{k}$ &  $\hat{o}$ &   $|E|$ &  $|V|$ &  \shortstack[l]{avarage \\ internal \\degree} &  \shortstack[l]{internal \\ edge \\density} &  \shortstack[l]{modularity \\ density} &  \shortstack[l]{normalized \\ cut} \\
    graph          &            &            &            &         &        &                                               &                                             &                                        &                                    \\
    \midrule
    PP\_0.20\_0.10   &       0.18 &       6.00 &       0.00 &   61.00 &  30.00 &                                          1.07 &                                        0.07 &                                  -2.32 &                               0.84 \\
    PP\_0.20\_0.15   &       0.18 &       6.00 &       0.00 &   84.00 &  30.00 &                                          0.93 &                                        0.06 &                                  -4.48 &                               0.96 \\
    PP\_0.20\_0.20   &       0.20 &       6.00 &       0.00 &   99.00 &  30.00 &                                          0.80 &                                        0.05 &                                  -6.00 &                               1.01 \\
    PP\_0.20\_0.25   &       0.31 &       6.00 &       0.00 &  119.00 &  30.00 &                                          0.80 &                                        0.05 &                                  -7.60 &                               1.04 \\
    PP\_0.40\_0.10   &       0.17 &       6.00 &       0.00 &   73.00 &  30.00 &                                          1.87 &                                        0.12 &                                  -1.36 &                               0.71 \\
\bottomrule
    \end{tabular}
}

\begin{flushleft}
  $|V|$ – the number of nodes; $|E|$ – the number of links; $\hat{k}$ – the number of known clusters; $\hat{o}$ – the number of overlapping nodes in the ground truth; $\hat{c}$ – the avg. clustering coefficient.
\end{flushleft}
\end{adjustwidth}
\end{table}

%%%%%%%%%%%%%%%%%%%%%%%%%%%%%%%%%%%%%%%%%%%%%%%%%%%%%%%%%%%%%%%%%%%%%%%%%%%%%%%%%%%%%%%%%%%%%%%%%%%%%%%%%%%%%%%%%%%%%%%%%%%%%%%%%%%%%%%%%%%%%
\subsection*{Implementation}

For historical reasons, we did two implementations of the LPAM method using Java and Python. Both can be found in our repository on GitHub at \url{https://github.com/aponom84/lpam-clustering}.
Java implementation includes exact and heuristic versions of the LPAM method. The Python version is available only for the heuristic.
The exact version solves the $k$-median problem, by employing an efficient mixed-integer linear programming model by Goldengorin~\cite{Goldengorin:2011}.
The Java heuristic implementations solves the $k$-median problem using an implementation of the CLARANS heuristic~\cite{ng2002clarans} from the {\tt smile} library~\cite{smile2}.
Python implementation commes as two methods ({\tt lpam\_python\_amp} for amplified commute distance, and {\tt lpam\_python\_cm} for commute distance) included into the script {\tt exp\_CDLIB.py}.
The script performs all computational experiments presented in the paper except experiments with the exact version.
The source code of the experiments with the exact version is in Jupyter notebooks which call Java-code.
\\The most of the code depends on the CDLIB \cite{giulio_rossetti_2021_4575156}, PyClustering \cite{Novikov2019}, and NetworkX \cite{hagberg2008exploring} libraries.

%%%%%%%%%%%%%%%%%%%%%%%%%%%%%%%%%%%%%%%%%%%%%%%%%%%%%%%%%%%%%%%%%%%%%%%%%%%%%%%%%%%%%%%%%%%%%%%%%%%%%%%%%%%%%%%%%%%%%%%%%%%%%%%%%%%%%%%%%%%%%%%%%%%%%%%
%%%%%%%%%%%%%%%%%%%%%%%%%%%%%%%%%%%%%%%%%%%%%%%%%%%%%%%%%%%%%%%%%%%%%%% Results %%%%%%%%%%%%%%%%%%%%%%%%%%%%%%%%%%%%%%%%%%%%%%%%%%%%%%%%%%%%%%%%%%%%%%%
\section*{Results}

In the table~\ref{tab:exact-onmi-table} we provide a comparision between the exact and heuristic versions of the LPAM method.
The OMNI values with an asterisk correspond to the cases where the heuristic solution of the $k$-median problem produces slightly better results compared to the the exact solution of the $k$-median problem.
This was caused by the fact that sometimes the ground-truth community structure does not match to neighbourhood of the medoids.
Thus, the mistake in the identification of the global minimum of the $k$-median problem leads to a solution which is closer to the ground truth.
% Moreover, the ground truth may not necessarily corresponds to the {\em true} community structure, given that we do not have the exact definition of a community in a network.

\begin{table}[!ht]
%  \begin{sidewaystable}[!htbp]
    \caption{\textbf{Comparison exact and heuristic versions of the LPAM method in tersm of ONMI}}\label{tab:exact-onmi-table}
    \centering
    %\begin{adjustwidth}{0in}{0in} % Comment out/remove adjustwidth environment if table fits in text column.
\resizebox{7cm}{!}{
      \begin{tabular}{ p{3.5cm} + p{1.7cm} | p{1.7cm} | p{1.7cm} | p{1.7cm}  }
        \hline
        \bf Name of network (Dataset) & \bf LPAM Exact CM & \bf LPAM Heuristic-CM & \bf LPAM Exact-ACM & \bf LPAM Heuristic-ACM
        \\
        \thickhline
    School Friendship & 0.555983 & $0.577798^{*}$ & \cellcolor{green!50} 0.748707 & \cellcolor{green!50} $0.748707 $ \\\hline
    Karate Club & \cellcolor{green!10}0.91796& \cellcolor{green!10} 0.91796 & \cellcolor{green!10} 0.91796 & \cellcolor{green!50}0.91796 \\ \hline
    Football league & 0.71194 & 0.71194 & - & \cellcolor{green!50} 0.916974 \\ \hline
    Adj-Noun & \cellcolor{green!10}0.00953164 & \cellcolor{green!10}0.00953164 & 0.00490969 & 0.00490969   \\ \hline
    Politics Book & 0.43517 & $0.440109^*$ & \cellcolor{green!50} 0.464154 & 0.422712 \\ \hline
    bench 30 & \cellcolor{green!50}0.931866 & \cellcolor{green!50}0.931866 & \cellcolor{green!50} 0.931866 & \cellcolor{green!50}0.931866 \\ \hline
    bench 40 & 0.203899 & 0.203899 & 0.346589 & 0.273879 \\ \hline
    bench 50 & 0.309034 & 0.275624 & \cellcolor{green!50} 0.845306 & \cellcolor{green!50} 0.845306 \\ \hline
    bench 60 & 0.405531 & 0.307907 & \cellcolor{green!10} 0.60211 & 0.586147 \\ \hline
    bench 60 dense & 0.112146 & 0.0671985 & 0.466922 & \cellcolor{green!10} 0.50446 \\ \hline

    %CKB-t $k=175, n=100-, \alpha=0.1, \gamma = 0.5$ max memb. = 20, max com,size=200 & - & - & - & 0.000784 & 0.0097659 & 0.0 & \cellcolor{green!50} 0.0110398\\ \hline
    \end{tabular}
    }
    \begin{flushleft}
      Each cell contains the ONMI values between the ground truth covering and the result of the corresponding method. The bright green cell background marks the highest (the best) ONMI value. Light green background means the second best result.
    \end{flushleft}
    %\end{adjustwidth}
    \end{table}
%    \end{sidewaystable}

    The computational times for the exact and the heuristic Java-versions of the LPAM method can be seen in Table~\ref{tab:time_efforts}.

    \begin{table}[!ht]
    \begin{adjustwidth}{0in}{0in}
    \caption{Time efforts for exact version of LPAM method in comparision with heuristic }\label{tab:time_efforts}
    %\centering
    \begin{tabular}{ p{1.1cm} + p{0.8cm} | p{0.7cm} | p{0.8cm} | p{1.0cm} | p{0.8cm} | p{0.8cm} | p{0.8cm} | p{0.8cm} | p{0.8cm} | p{0.8cm} | p{0.8cm} }
    \hline
    \bf Dataset Name & School & Karate & Football & Adj-Noun&Politics Book&bench 30&bench 40& bench 50&bench 60&dense 60\\
    \hline
    \bf $|V|$ & 69 & 34 & 62 & 112 & 105 & 30 & 40 & 50 & 60 & 60\\
    \hline
    \bf $|E|$ & 194 & 78 & 613 & 425 & 441 & 120 & 220 & 282 & 260 & 262 \\

    \thickhline
    Exact & 29.938 & 1.542 & - &1448 & 817 & 0.885 & 6.082 & 11.617 & 5.025 & 7.342\\ \hline
    $\theta$ & 0.55 & 0.45 & - & 0.1 & 0.55 & 0.35 & 0.35 & 0.35 & 0.55 & 0.2 \\
    \thickhline
    Heuristic & 0.397 & 0.227 & 0.712 & 0.553 & 0.594 & 0.179 & 0.388 & 0.548 & 0.336 & 0.422 \\
    \hline
    $\theta$ & 0.55 & 0.45 & 0.55 & 0.1 & 0.55 & 0.35 & 0.05 & 0.35 & 0.35 & 0.35 \\
    \thickhline
    Matrix calculation & 3.337 & 0.194 & 832 & 97 & 143 & 0.144 & 0.393 & 0.922 & 0.683 & 0.659\\
    \hline\end{tabular}
    \begin{flushleft}
    The time consumptions of the LPAM method (exact and heuristic versions) in seconds. The rows with $\theta$ contain optimal values of the threshold parameter.
    Computations were done on an Intel Xeon with X5675 3.07GHz. IBM ILOG CPLEX Optimizer 12.9.0 was used as the integer programming solver.
    \end{flushleft}
    \end{adjustwidth}

    \end{table}

We studied median and maximum values for the F1, omega index, and overlapping normalized mutual information (ONMI) as scoring functions for measuring similarities between the partitioning produced by a method and the ground truth..
Also we studied intra cluster edge density, and normalized cut \cite{shi2000normalized}.
As can been seen from the table~\ref{tab:f1_results_median}, the LPAM method gives the best median value of F1 score for several
instances of the planted partition model. The GCE and Leiden methods have the best results in most cases.
However, results for maximum F1 score \ref{tab:f1_results_median}, maximum omega index \ref{tab:omega_results_max},
and maximum ONMI \ref{tab:onmi_results_max} show that the LPAM method equpited by amplified commute distance gives the best scores for the most PP-graphs,
football club, and lattice8x8.
Interesting, that LPAM method with commute distance able to find the exact solution (F1, omega index, and ONMI scores are 1) for the karate club.

In terms of internal cluster density tables \ref{tab:edge_density_results_median}, \ref{tab:edge_density_results_max}, show that the best solutions can be found by the LPAM-amp method for PP/SBM-graph.
The commute distance with the LPAM method gives most density partitioning for FARZ networks,
while the k-clique method can produce most density solutions for Lancichinetti and FARZ networks.

Partitioning where no nodes belonging to any community has a zero value of normalized cut. That is why we have many zero values in the
tables~\ref{tab:normalized_cut_results_median}, \ref{tab:normalized_cut_results_min}.

Generally, based on our communicational experiments we can summarize that there is no method that dominates the rest with respect to the proximity to the ground-truth covering for all data sets.

The average, and maximum execution times for each compared clustering methods are presented in table~\ref{tab:time_results_average}, and table~\ref{tab:time_results_max} respectively.

\begin{figure}[h]
  \centering
  \includegraphics[scale=0.6]{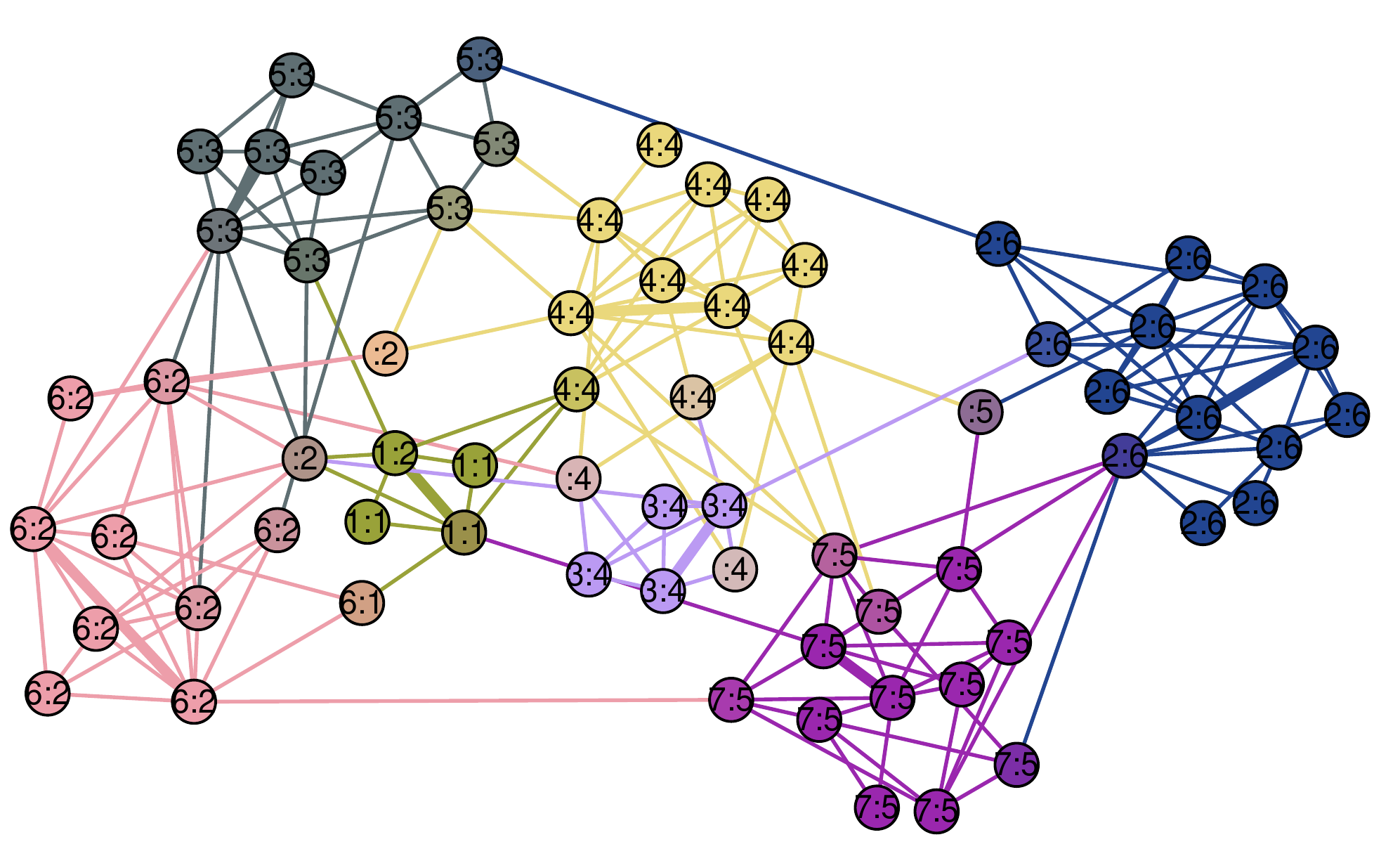}
\caption{Clustering results of the exact version of the LPAM method with amplified commute distance for the {\tt School Friendship} network.
$\theta = 0.5, k=7$. The paired numbers, separated by a colon inside the nodes, denote the predicted ID of the cluster provided by the LPAM method and the ground truth, respectively.
In addition, the LPAM algorithm covering are denoted by colors.
As seen in the picture, the algorithm correctly revealed community 3, 5, and 6.
Also, the algorithm does not assign the nodes that have connections with more than two clusters, to any cluster. That is because of the parameter $\theta = 0.5$
The LPAM method falsely separates community 4 into two clusters (4 - yellow and 3 - purple).
Moreover, the algorithm almost correctly identifies the small community 1 (green); however, it incorrectly assigns the node from community 2 to this cluster.}
\label{fig:school-lpam-acm}
\end{figure}

\begin{figure}[h]
  \centering
  \includegraphics[scale=0.6]{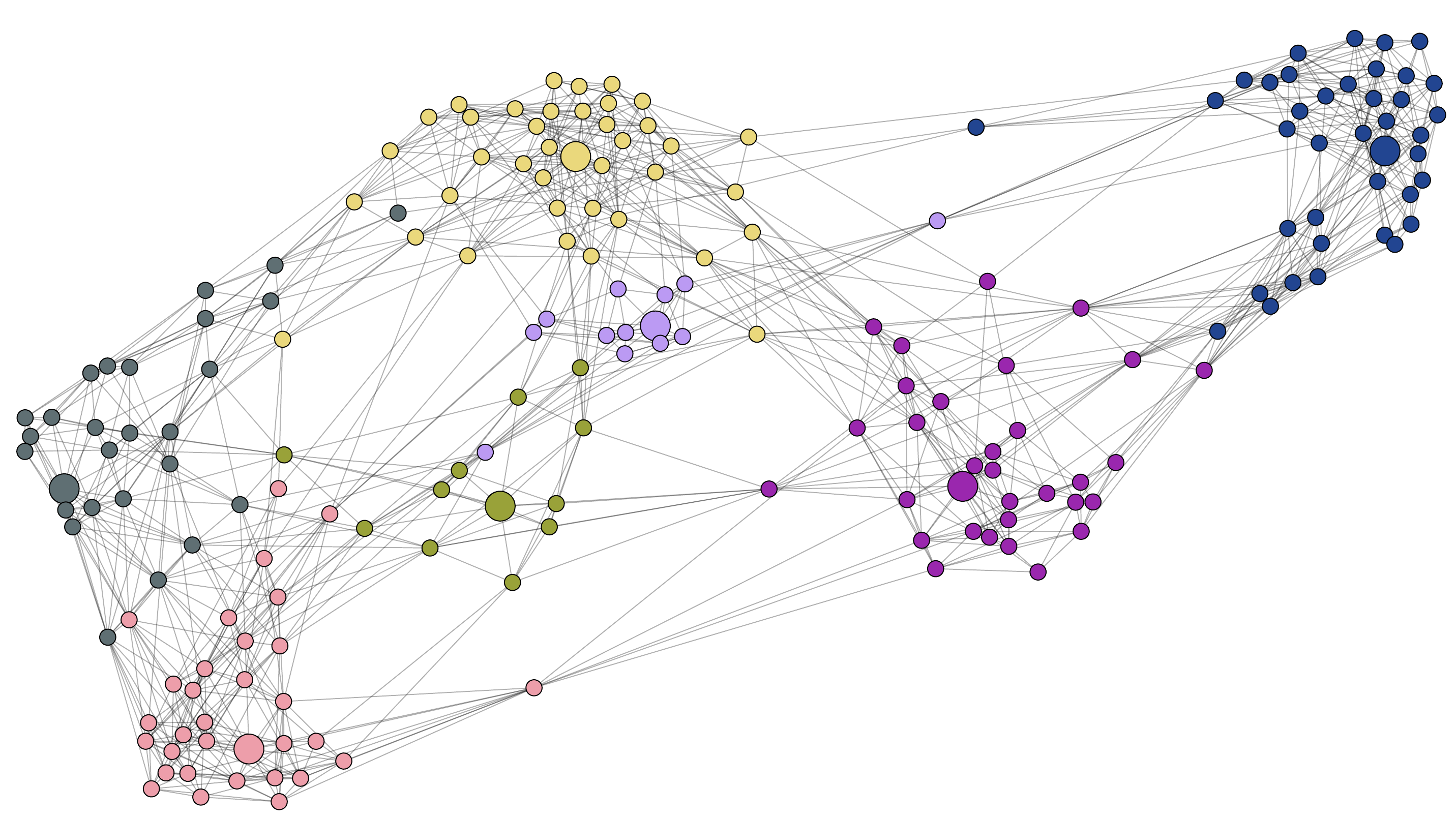}
\caption{A line graph which is produced by the exact version of the LPAM method with amplified commute distance for the {\tt School Friendship} network ($\theta = 0.5, k= 7$). }\label{fig:school-lpam-acm-line}
\end{figure}

An example of method output for the {\tt School Friendship} instance can be seen in the Figure~\ref{fig:school-lpam-acm} and the associated line graph in Figure~\ref{fig:school-lpam-acm-line}.

We should also note that the LPAM method for the regular lattice example naturally produces the covering that matches the natural separation, in contrast to the GCE method which does not produce any result for this example, while the OSLOM method can only give an accidentally good result with appropriately chosen input parameters.

% \end{sidewaystable}

\begin{table}
  \caption{F1 median score results table. Maximum values are highlighted by green backgound}
  \label{tab:f1_results_median}
  \resizebox{7cm}{!}{
\begin{tabular}{llllllllllllllll}
      \toprule
      method &                        GCE & \shortstack[l]{egonet \\ splitter} & \shortstack[l]{label \\ propagation} & \shortstack[l]{lpam \\ amp} & \shortstack[l]{lpam \\ cm} & angel &                      copra &                      danmf & demon &                    kclique & lais2 &                     leiden &                      oslom &                   percomvc &  slpa \\
      graph        &                            &                                    &                                      &                             &                            &       &                            &                            &       &                            &       &                            &                            &                            &       \\
      \midrule
      PP\_0.20\_0.10 &                       0.29 &                               0.36 &                                 0.30 &                        0.37 &                       0.39 &  0.35 &                       0.29 &                       0.34 &  0.35 &                       0.37 &  0.40 &  \cellcolor{green!50} 0.43 &                       0.30 &                       0.29 &  0.32 \\
      PP\_0.20\_0.15 &                       0.29 &                               0.33 &                                 0.38 &   \cellcolor{green!50} 0.42 &                       0.36 &  0.36 &                       0.29 &                       0.33 &  0.34 &                       0.33 &  0.31 &                       0.37 &                       0.29 &                       0.29 &  0.29 \\
\bottomrule
      \end{tabular}

    }
\end{table}

\begin{table}
  \caption{F1 max score results table. Maximum values are highlighted by green backgound}
  \label{tab:f1_results_max}
  \resizebox{7cm}{!}{
    \begin{tabular}{llllllllllllllll}
      \toprule
      method &                        GCE & \shortstack[l]{egonet \\ splitter} & \shortstack[l]{label \\ propagation} & \shortstack[l]{lpam \\ amp} & \shortstack[l]{lpam \\ cm} & angel &                      copra &                      danmf &                      demon &                    kclique & lais2 &                     leiden &                      oslom &                   percomvc &                       slpa \\
      graph        &                            &                                    &                                      &                             &                            &       &                            &                            &                            &                            &       &                            &                            &                            &                            \\
      \midrule
      PP\_0.20\_0.10 &                       0.29 &                               0.36 &                                 0.30 &                        0.43 &                       0.42 &  0.35 &                       0.32 &                       0.45 &                       0.36 &                       0.45 &  0.40 &                       0.43 &  \cellcolor{green!50} 0.48 &                       0.29 &                       0.40 \\
      PP\_0.20\_0.15 &                       0.29 &                               0.33 &                                 0.38 &   \cellcolor{green!50} 0.49 &                       0.48 &  0.37 &                       0.30 &                       0.45 &                       0.43 &                       0.37 &  0.31 &                       0.37 &                       0.29 &                       0.29 &                       0.37 \\
\bottomrule
      \end{tabular}
    }
\end{table}

\section*{Discussion}\label{sec:Discussion}
%%%%%%%%%%%%%%%%%%%%%%%%%%%%%%%%%%%%%%%%%%%%%%%%%%%%%%%%%%%%%%%%%%%%%%%%%%%%%%%%%%%%%%%%%%%%%%%%%%%%%%%%%%%%%%%%%%%%%%%%%%%%%%%%%%%%%%%%%%%%%
\subsection*{School Friendship example}
For the school friendship network, the LPAM method produced quite accurate results.

In figure \ref{fig:school-lpam-acm}, the pairs of numbers, separated by a colon inside the nodes, denote the predicted ID of the cluster provided by the LPAM method and the ground truth, respectively.
In addition, the LPAM algorithm coverings are denoted by colors.
As seen in the picture, the algorithm correctly finds communities 3, 5, and 6.
The two nodes that belong to 3 clusters also have been detected correctly.
LPAM method wrongly separates community 4 into the two clusters (yellow and purple).
Also, it wrongly decides that one node from community 2 belongs to the green cluster, but the algorithm almost correctly identified the small community 1.
Lastly, it should be noted that the community structure produced by LPAM algorithm for 7 clusters ($k=7$) has a larger value of ONMI score than for 6 clusters ($k=6$).

\begin{figure}[h]
\centering \includegraphics[width=8cm]{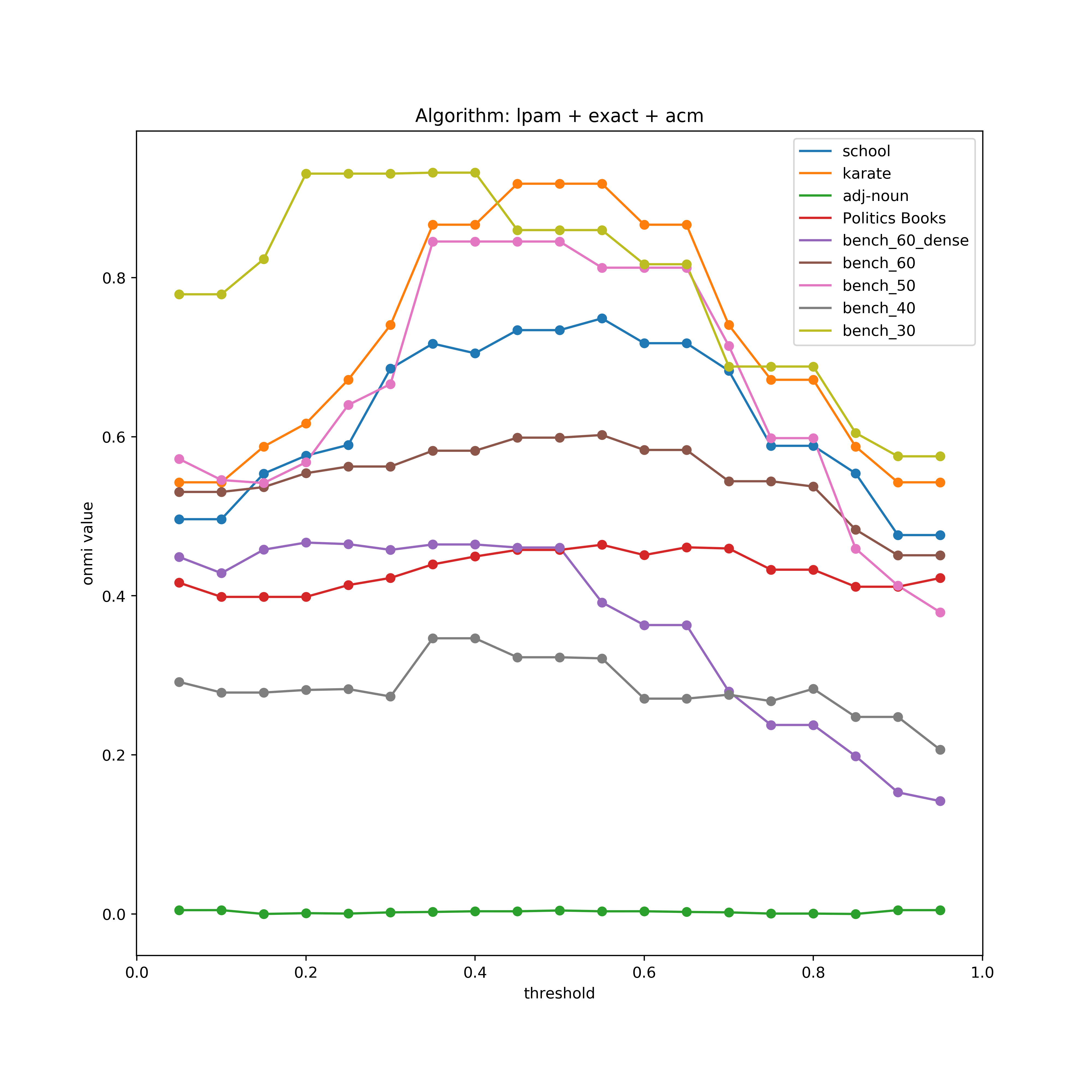}
\caption{The ONMI results for the exact version of the LPAM method with amplified commute distance depending on the threshold parameter $\theta$.}
\label{fig:lpam_exact_onmi}
\end{figure}

%%%%%%%%%%%%%%%%%%%%%%%%%%%%%%%%%%%%%%%%%%%%%%%%%%%%%%%%%%%%%%%%%%%%%%%%%%%%%%%%%%%%%%%%%%%%%%%%%%%%%%%%%%%%%%%%%%%%%%%%%%%%%%%%%%%%%%%%%%%%%
\subsection*{Tuning parameters}The behaviour of the ONMI value depending on the threshold parameter $\theta$ for the LPAM method is shown in Figure~\ref{fig:lpam_exact_onmi}. As it can be seen in most cases the maximum ONMI value is reached when the threshold value $\theta$ lies between $0.3$ and $0.6$. This can be attributed to the fact that for the proximity to the ground truth covering, it is usually better to either assign a vertex to one cluster or no cluster rather than to several clusters.
The clustering results of the heuristic version of the LPAM method with amplified commute distance for the FARZ networks with 200 nodes and 5 communities can be found in Figure~\ref{tab:FARZ}.

%The final coverings with the four various combination of the exact/heuristic version of the LPAM method, with commute and amplified commute distances for all datasets, are presented in the appendices \nameref{S1_lpam_amp_exact},\nameref{S2_lpam_amp_heuristic}, \nameref{S3_lpam_cm_exact}, and \nameref{S4_lpam_cm_heuristic}.
%The resulting pictures for the best ONMI values, as well as the full study of the dependence of the ONMI value on the input parameters for the GCE, OSLOM, and COPRA methods, can be found in the appendices \nameref{S5_GCE}, \nameref{S6_OSLOM}, and \nameref{S7_COPRA}, respectively.

%\begin{figure}[h]
%  \centering
%%\includegraphics[width=8cm]{figures/lpam_exact_acm_onmi.png}
%\caption{The clustering results of the heuristic version of the LPAM method with amplified commute distance for the FARZ networks with 200 nodes and 5 communities. (a) $\beta = 1$, (b) $\beta = 0.95$, (c) $\beta = 0.9 $, (d) $\beta = 0.85$, (e) $\beta = 0.8$, (f) $\beta = 0.75 $, (g) $\beta = 0.7$, (h) $\beta = 0.65$, (i) $\beta = 0.6 $, (j) $\beta = 0.55$, (k) $\beta = 0.5$. }
%\label{fig:FARZ}
%\end{figure}

%%%%%%%%%%%%%%%%%%%%%%%%%%%%%%%%%%%%%%%%%%%%%%%%%%%%%%%%%%%%%%%%%%%%%%%%%%%%%%%%%%%%%%%%%%%%%%%%%%%%%%%%%%%%%%%%%%%%%%%%%%%%%%%%%%%%%%%%%%%%%
\subsection*{Computational Complexity} \label{subsec:computational_complexity}

The computation complexity of the LPAM method consists of three parts: TIME(building line graph) + TIME(calculating distance matrix) + TIME(solving $k$-median problem).
For the first term TIME(building line graph), we need $\Theta(|E|)$ time.
The second term TIME(calculating distance matrix) depends on the distance type used.
Thus, a matrix with the shortest path distances can be calculated with the Floyd--Warshall algorithm~\cite{floyd1962algorithm} with cubic time on the number of nodes of the linear graph, which means $\Theta(|E|^3)$ time. Both commute distance and amplified commute distance require the calculation of the Moore-–Penrose pseudo-inverse matrix.
Naively, we need $\Theta(|E|^3)$ time to calculate the Moore-–Penrose pseudo-inverse matrix because we need to get $|E|$ eigenvectors.
However, theoretically, it is possible to find an algorithm with computational complexity close to quadratic~\cite{demmel2007fast}.

\subsubsection*{Exact version}
Because the $k$-median problem is NP-complete, it has exponential complexity.
In the case of link partitioning the complexity is $O(2^{|E|})$.
Therefore, the total complexity of the LPAM method is dominated by the third exponential term TIME(solving $k$-median problem).
\subsubsection*{Heuristic version}
In the case of the heuristic, to obtain a solution we used the CLARANS method which can be considered as a kind of randomized local search heuristic or variable neighbourhood search.
CLARANS is an iterative heuristic; on each iteration, it tries to improve the current solution and stops when this is not feasible.
It is hard to determine the number of iterations the algorithm makes until it stops.
The rough upper bound is the size of the solution space, which, in turn, is exponential.
Moreover, the CLARANS method makes several attempts to obtain various local minima and stops when no improvement is made.
Thus, it is hard to establish the computational complexity bounds. Like many iterative heuristics, CLARANS stops when it cannot improve the solution.
%Thus it makes it hard to build any bounds of computational complexity. Moreover, there is a trade-off between computational efforts and accuracy. The trade-off is tuned by the number of attempts (iteration) to obtain local minimums.

%%%%%%%%%%%%%%%%%%%%%%%%%%%%%%%%%%%%%%%%%%%%%%%%%%%%%%%%%%%%%%%%%%%%%%%%%%%%%%%%%%%%%%%%%%%%%%%%%%%%%%%%%%%%%%%%%%%%%%%%%%%%%%%%%%%%%%%%%%%%%%% Discussion%%%%%%%%%%%%%%%%%%%%%%%%%%%%%%%%%%%%%%%%%%%%%%%%%%%%%%%%%%%%%%%%%%%%%%%%%%%%%%%%%%%%%%%%%%%%%%%%%%%%%%%%%%%%%%%%%%%%%%%%%%%%%%%%%%%%%%%%%%%%%

%%%%%%%%%%%%%%%%%%%%%%%%%%%%%%%%%%%%%%%%%%%%%%%%%%%%%%%%%%%%%%%%%%%%%%%%%%%%%%%%%%%%%%%%%%%%%%%%%%%%%%%%%%%%%%%%%%%%%%%%%%%%%%%%%%%%%%%%%%%%%
%%  Discussion
%%%%%%%%%%%%%%%%%%%%%%%%%%%%%%%%%%%%%%%%%%%%%%%%%%%%%%%%%%%%%%%%%%%%%%%%%%%%%%%%%%%%%%%%%%%%%%%%%%%%%%%%%%%%%%%%%%%%%%%%%%%%%%%%%%%%%%%%%%%%%

%%%%%%%%%%%%%%%%%%%%%%%%%%%%%%%%%%%%%%%%%%%%%%%%%%%%%%%%%%%%%%%%%%%%%%%%%%%%%%%%%%%%%%%%%%%%%%%%%%%%%%%%%%%%%%%%%%%%%%%%%%%%%%%%%%%%%%%%%%%%%
%%  Conclusion
%%%%%%%%%%%%%%%%%%%%%%%%%%%%%%%%%%%%%%%%%%%%%%%%%%%%%%%%%%%%%%%%%%%%%%%%%%%%%%%%%%%%%%%%%%%%%%%%%%%%%%%%%%%%%%%%%%%%%%%%%%%%%%%%%%%%%%%%%%%%%

\section*{Conclusion}\label{sec:Conclusion}
In this paper, we propose a new method for the detection of overlapping communities in networks with a predefined number of clusters.
The proposed method is based on finding disjoint communities on the line graph of the original network, by partitioning around medoids.
The resulting link partitioning naturally produces an overlapping community structure for the original graph.
The link partitioning uses commute distance and its variation that produces more accurate results.

Experimental results on the set of well-known benchmark instances, as well as artificially generated instances with known ground truth, demonstrate that the proposed method can compete with the existing methods in the literature, which motivates us to further improve the method.
The computation results demonstrated that the heuristic version produces results that are very close to the exact version.

\section*{Acknowledgments}
The article was prepared within the framework of the Basic Research Program at the National Research University Higher School of Economics.\\
The authors would like to thank Anna Yaushkina and Nikita Putikhin for their help with the implementation of the amplified commute distance function, and for implementing the heuristic Java version of the LPAM method. A special thanks goes to Eldar Yusupov who set up the computational cluster of the LATNA laboratory at the HSE. We also thank Alexey Malafeev who helped to improve the language of the paper.
Also, authors would like to thank Peter Miasnikof who found a bug in the python implementation of the amplified commute distance.
%\nolinenumbers

%%%%%%%%%%%%%%%%%%%%%%%%%%%%%%%%%%%%%%%%%%%%%%%%%%%%%%%%%%%%%%%%%%%%%%%%%%%%%%%%%%%%%%%%%%%%%%%%%%%%%%%%%%%%%%%%%%%%%%%%%%%%%%%%%%%%%%%%%%%%%
%%  Bibliography
%%%%%%%%%%%%%%%%%%%%%%%%%%%%%%%%%%%%%%%%%%%%%%%%%%%%%%%%%%%%%%%%%%%%%%%%%%%%%%%%%%%%%%%%%%%%%%%%%%%%%%%%%%%%%%%%%%%%%%%%%%%%%%%%%%%%%%%%%%%%%

%\bibliographystyle{plos2015}
\bibliographystyle{unsrt}
\bibliography{lpam_arxiv.bib}

\section*{Appendix: Extended Results}

\begin{table}[ht]
\begin{adjustwidth}{0in}{0in} % Comment out/remove adjustwidth environment if table fits in text column.
\caption{Clustering results of heuristic version of LPAM method with Amplified Commute Distance for FARZ networks with 200 nodes and 20 communities}

\centering
%
\label{tab:FARZ}
\end{adjustwidth}
\end{table}

\begin{figure}[h]
\centering
\includegraphics[scale=0.8]{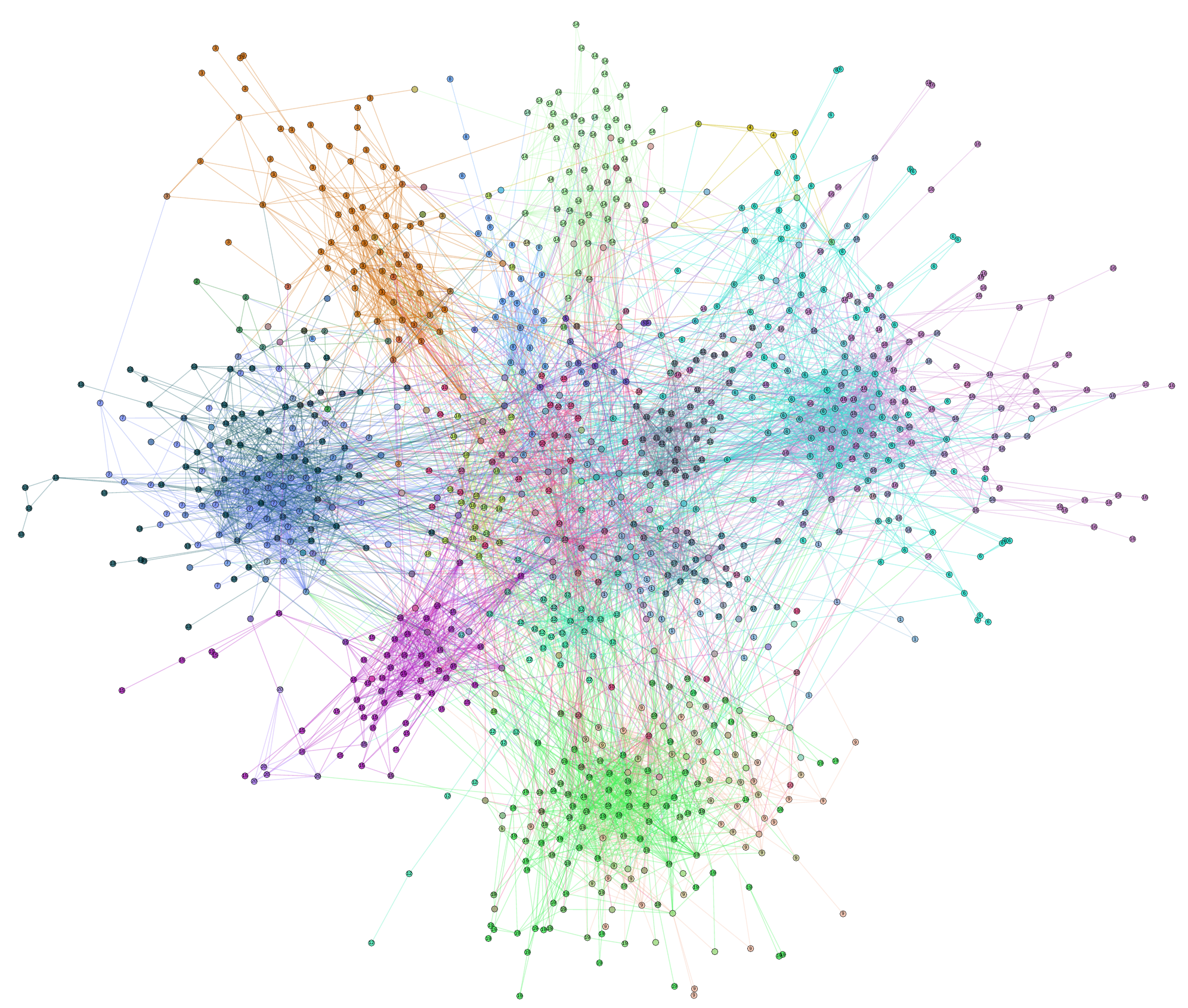} \

\caption{Clustering result of LPAM method with amplified commute distance function on FARZ network $n=1000, m=7, k = 20, \beta = 0.75$} 
\end{figure}

\begin{table}
  \caption{Median values of omega index. Maximum values are highlighted by green backgound}
  \label{tab:omega_results_median}
  \resizebox{10cm}{!}{
    \begin{tabular}{llllllllllllllll}
      \toprule
      method &   GCE & \shortstack[l]{egonet \\ splitter} & \shortstack[l]{label \\ propagation} & \shortstack[l]{lpam \\ amp} & \shortstack[l]{lpam \\ cm} &                      angel &                      copra &                      danmf &                      demon &                    kclique &                      lais2 &                     leiden &                      oslom &                   percomvc &                       slpa \\
      graph        &       &                                    &                                      &                             &                            &                            &                            &                            &                            &                            &                            &                            &                            &                            &                            \\
      \midrule
      PP\_0.20\_0.10 &  0.00 &          \cellcolor{green!50} 0.07 &                                 0.01 &                        0.04 &                       0.04 &                       0.04 &                       0.00 &                      -0.01 &                       0.04 &                       0.02 &                       0.05 &                       0.07 &                       0.01 &                       0.00 &                       0.01 \\
      PP\_0.20\_0.15 &  0.00 &                               0.01 &                                -0.00 &                        0.06 &                       0.02 &                       0.02 &                       0.00 &                      -0.00 &  \cellcolor{green!50} 0.09 &                       0.03 &                       0.01 &                       0.04 &                       0.00 &                       0.00 &                      -0.01 \\
\bottomrule
      \end{tabular}
    }
\end{table}

\begin{table}
  \caption{Maximum values of omega index for each clustering method. Maximum values are highlighted by green backgound}
  \label{tab:omega_results_max}
  \resizebox{10cm}{!}{
    \begin{tabular}{llllllllllllllll}
      \toprule
      method &                        GCE & \shortstack[l]{egonet \\ splitter} & \shortstack[l]{label \\ propagation} & \shortstack[l]{lpam \\ amp} & \shortstack[l]{lpam \\ cm} &                      angel &                      copra &                      danmf &                      demon &                    kclique &  lais2 &                     leiden &                      oslom &                   percomvc &                       slpa \\
      graph        &                            &                                    &                                      &                             &                            &                            &                            &                            &                            &                            &        &                            &                            &                            &                            \\
      \midrule
      PP\_0.20\_0.10 &                       0.00 &                               0.07 &                                 0.01 &   \cellcolor{green!50} 0.10 &                       0.10 &                       0.04 &                       0.00 &                       0.05 &                       0.04 &                       0.05 &   0.05 &                       0.07 &                       0.09 &                       0.00 &                       0.07 \\
      PP\_0.20\_0.15 &                       0.00 &                               0.01 &                                -0.00 &                        0.12 &  \cellcolor{green!50} 0.14 &                       0.02 &                       0.00 &                       0.09 &                       0.13 &                       0.05 &   0.01 &                       0.04 &                       0.00 &                       0.00 &                       0.04 \\
\bottomrule
      \end{tabular}

    }
\end{table}

\begin{table}
  \caption{Median values of overlapping normalized mutual information (LFK). Maximum values are highlighted by green backgound}
  \label{tab:onmi_results_median}
  \resizebox{10cm}{!}{
    \begin{tabular}{llllllllllllllll}
      \toprule
      method &   GCE & \shortstack[l]{egonet \\ splitter} & \shortstack[l]{label \\ propagation} & \shortstack[l]{lpam \\ amp} & \shortstack[l]{lpam \\ cm} &                      angel &                      copra &                      danmf &                      demon &                    kclique &                      lais2 &                     leiden &                      oslom &                   percomvc &                       slpa \\
      graph        &       &                                    &                                      &                             &                            &                            &                            &                            &                            &                            &                            &                            &                            &                            &                            \\
      \midrule
      PP\_0.20\_0.10 &  0.00 &          \cellcolor{green!50} 0.07 &                                 0.01 &                        0.04 &                       0.04 &                       0.04 &                       0.00 &                      -0.01 &                       0.04 &                       0.02 &                       0.05 &                       0.07 &                       0.01 &                       0.00 &                       0.01 \\
      PP\_0.20\_0.15 &  0.00 &                               0.01 &                                -0.00 &                        0.06 &                       0.02 &                       0.02 &                       0.00 &                      -0.00 &  \cellcolor{green!50} 0.09 &                       0.03 &                       0.01 &                       0.04 &                       0.00 &                       0.00 &                      -0.01 \\
\bottomrule
      \end{tabular}

    }
\end{table}

\begin{table}
  \caption{Maximum values of overlapping normalized mutual information (LFK) producced by clustering algorithms. Maximum values in a row are highlighted by green backgound}
  \label{tab:onmi_results_max}
  \resizebox{10cm}{!}{
    \begin{tabular}{llllllllllllllll}
      \toprule
      method &                        GCE & \shortstack[l]{egonet \\ splitter} & \shortstack[l]{label \\ propagation} & \shortstack[l]{lpam \\ amp} & \shortstack[l]{lpam \\ cm} & angel &                      copra &                      danmf &                      demon &                    kclique & lais2 &                     leiden &                      oslom & percomvc &                       slpa \\
      graph        &                            &                                    &                                      &                             &                            &       &                            &                            &                            &                            &       &                            &                            &          &                            \\
      \midrule
      PP\_0.20\_0.10 &                       0.00 &                               0.08 &                                 0.20 &                        0.16 &  \cellcolor{green!50} 0.23 &  0.12 &                       0.07 &                       0.15 &                       0.12 &                       0.12 &  0.12 &                       0.07 &                       0.20 &     0.00 &                       0.19 \\
      PP\_0.20\_0.15 &                       0.00 &                               0.04 &                                 0.07 &                        0.17 &  \cellcolor{green!50} 0.24 &  0.05 &                       0.02 &                       0.15 &                       0.18 &                       0.07 &  0.06 &                       0.07 &                       0.00 &     0.00 &                       0.15 \\
\bottomrule
      \end{tabular}

    }
\end{table}

\begin{table}
  \caption{Median values of internal cluster edge density. Maximum values are highlighted by green backgound}
  \label{tab:edge_density_results_median}
  \resizebox{10cm}{!}{
    \begin{tabular}{llllllllllllllll}
      \toprule
      method &   GCE & \shortstack[l]{egonet \\ splitter} & \shortstack[l]{label \\ propagation} & \shortstack[l]{lpam \\ amp} & \shortstack[l]{lpam \\ cm} & angel & copra & danmf & demon &                    kclique &                      lais2 &                     leiden &                      oslom &                   percomvc &                       slpa \\
      graph        &       &                                    &                                      &                             &                            &       &       &       &       &                            &                            &                            &                            &                            &                            \\
      \midrule
      PP\_0.20\_0.10 &  0.04 &                               0.05 &            \cellcolor{green!50} 0.19 &                        0.14 &                       0.15 &  0.15 &  0.04 &  0.03 &  0.15 &                       0.12 &                       0.09 &                       0.13 &                       0.11 &                       0.04 &                       0.14 \\
      PP\_0.20\_0.15 &  0.05 &                               0.05 &                                 0.12 &                        0.13 &  \cellcolor{green!50} 0.13 &  0.12 &  0.05 &  0.02 &  0.11 &                       0.09 &                       0.06 &                       0.12 &                       0.05 &                       0.05 &                       0.09 \\
      PP\_0.20\_0.20 &  0.06 &                               0.11 &                                 0.08 &   \cellcolor{green!50} 0.15 &                       0.10 &  0.06 &  0.06 &  0.02 &  0.08 &                       0.13 &                       0.14 &                       0.14 &                       0.06 &                       0.09 &                       0.06 \\
\bottomrule
      \end{tabular}
    }
\end{table}

\begin{table}
  \caption{Maximux Internal cluster edge density median. Maximum values are highlighted by green backgound}
  \label{tab:edge_density_results_max}
  \resizebox{10cm}{!}{
    \begin{tabular}{llllllllllllllll}
      \toprule
      method &   GCE & \shortstack[l]{egonet \\ splitter} & \shortstack[l]{label \\ propagation} & \shortstack[l]{lpam \\ amp} & \shortstack[l]{lpam \\ cm} & angel &                      copra &                      danmf & demon &                    kclique &                      lais2 &                     leiden &                      oslom & percomvc &                       slpa \\
      graph        &       &                                    &                                      &                             &                            &       &                            &                            &       &                            &                            &                            &                            &          &                            \\
      \midrule
      PP\_0.20\_0.10 &  0.04 &                               0.05 &                                 0.19 &                        0.17 &                       0.17 &  0.15 &                       0.04 &                       0.06 &  0.17 &  \cellcolor{green!50} 0.21 &                       0.09 &                       0.13 &                       0.18 &     0.04 &                       0.16 \\
      PP\_0.20\_0.15 &  0.05 &                               0.05 &                                 0.12 &   \cellcolor{green!50} 0.18 &                       0.17 &  0.12 &                       0.05 &                       0.07 &  0.17 &                       0.14 &                       0.06 &                       0.12 &                       0.05 &     0.05 &                       0.14 \\
\bottomrule
      \end{tabular}
    }
\end{table}

\begin{table}
  \caption{Median values of normalized cut. Maximum values are highlighted by green backgound}
  \label{tab:normalized_cut_results_median}
  \resizebox{10cm}{!}{
    \begin{tabular}{llllllllllllllll}
      \toprule
      method &   GCE & \shortstack[l]{egonet \\ splitter} & \shortstack[l]{label \\ propagation} & \shortstack[l]{lpam \\ amp} & \shortstack[l]{lpam \\ cm} & angel & copra &                      danmf & demon & kclique &                      lais2 &                     leiden &                      oslom & percomvc &  slpa \\
      graph        &       &                                    &                                      &                             &                            &       &       &                            &       &         &                            &                            &                            &          &       \\
      \midrule
      PP\_0.20\_0.10 &  0.00 &                               0.98 &                                 0.48 &                        0.55 &                       0.54 &  0.52 &  0.00 &  \cellcolor{green!50} 1.00 &  0.52 &    0.30 &                       0.63 &                       0.47 &                       0.46 &     0.00 &  0.69 \\
      PP\_0.20\_0.15 &  0.00 &                               1.00 &                                 0.57 &                        0.71 &                       0.74 &  0.74 &  0.00 &  \cellcolor{green!50} 1.05 &  0.82 &    0.38 &                       0.66 &                       0.54 &                       0.00 &     0.00 &  0.69 \\
      PP\_0.20\_0.20 &  0.00 &                               0.97 &                                 0.51 &                        0.77 &                       0.76 &  0.62 &  0.00 &  \cellcolor{green!50} 1.08 &  0.84 &    0.70 &                       0.70 &                       0.64 &                       0.00 &     0.81 &  0.66 \\
\bottomrule
      \end{tabular}
    }
\end{table}

\begin{table}
  \caption{Minimum values of normalized cut produced by a method. Minimum values are highlighted by green backgound}
  \label{tab:normalized_cut_results_min}
  \resizebox{10cm}{!}{
    \begin{tabular}{llllllllllllllll}
      \toprule
      method &                        GCE & \shortstack[l]{egonet \\ splitter} & \shortstack[l]{label \\ propagation} & \shortstack[l]{lpam \\ amp} & \shortstack[l]{lpam \\ cm} &                      angel &                      copra & danmf &                      demon &                    kclique &                      lais2 & leiden &                      oslom &                   percomvc &                       slpa \\
      graph        &                            &                                    &                                      &                             &                            &                            &                            &       &                            &                            &                            &        &                            &                            &                            \\
      \midrule
      PP\_0.20\_0.10 &  \cellcolor{green!50} 0.00 &                               0.98 &                                 0.48 &                        0.38 &                       0.42 &                       0.52 &  \cellcolor{green!50} 0.00 &  0.91 &                       0.52 &  \cellcolor{green!50} 0.00 &                       0.63 &   0.47 &  \cellcolor{green!50} 0.00 &  \cellcolor{green!50} 0.00 &                       0.49 \\
      PP\_0.20\_0.15 &  \cellcolor{green!50} 0.00 &                               1.00 &                                 0.57 &                        0.52 &                       0.50 &                       0.74 &  \cellcolor{green!50} 0.00 &  1.00 &                       0.74 &  \cellcolor{green!50} 0.00 &                       0.66 &   0.54 &  \cellcolor{green!50} 0.00 &  \cellcolor{green!50} 0.00 &  \cellcolor{green!50} 0.00 \\
\bottomrule
      \end{tabular}

    }
\end{table}

\begin{table}
  \caption{The average execution time(seconds). Minimum values are highlighted by green backgound}
  \label{tab:time_results_average}
  \resizebox{10cm}{!}{
    \begin{tabular}{llllllllllllllll}
      \toprule
      method &   GCE & \shortstack[l]{egonet \\ splitter} & \shortstack[l]{label \\ propagation} & \shortstack[l]{lpam \\ amp} & \shortstack[l]{lpam \\ cm} & angel & copra & danmf & demon &                    kclique &   lais2 &                     leiden & oslom &                   percomvc &  slpa \\
      graph        &       &                                    &                                      &                             &                            &       &       &       &       &                            &         &                            &       &                            &       \\
      \midrule
      PP\_0.20\_0.10 &  0.14 &                               0.02 &                                 0.00 &                        0.00 &                       0.00 &  0.01 &  0.81 &  0.15 &  0.03 &                       0.00 &    0.54 &                       0.00 &  0.39 &  \cellcolor{green!50} 0.00 &  0.26 \\
      PP\_0.20\_0.15 &  0.14 &                               0.03 &                                 0.00 &                        0.00 &                       0.00 &  0.02 &  0.80 &  0.17 &  0.03 &                       0.00 &    0.57 &                       0.00 &  0.38 &  \cellcolor{green!50} 0.00 &  0.33 \\
      PP\_0.20\_0.20 &  0.14 &                               0.04 &                                 0.00 &                        0.00 &                       0.00 &  0.02 &  0.80 &  0.17 &  0.04 &  \cellcolor{green!50} 0.00 &    0.58 &                       0.00 &  0.40 &                       0.00 &  0.39 \\
      PP\_0.20\_0.25 &  0.14 &                               0.03 &                                 0.00 &                        0.00 &  \cellcolor{green!50} 0.00 &  0.02 &  0.78 &  0.17 &  0.06 &                       0.00 &    0.48 &                       0.00 &  0.42 &                       0.01 &  0.46 \\
\bottomrule
      \end{tabular}
    }
\end{table}

\begin{table}
  \caption{Maximum values of time execution (seconds). Minimum values are highlighted by green backgound}
  \label{tab:time_results_max}
  \resizebox{10cm}{!}{
    \begin{tabular}{llllllllllllllll}
      \toprule
      method &   GCE & \shortstack[l]{egonet \\ splitter} & \shortstack[l]{label \\ propagation} & \shortstack[l]{lpam \\ amp} & \shortstack[l]{lpam \\ cm} & angel & copra & danmf & demon &                    kclique &   lais2 &                     leiden & oslom &                   percomvc &  slpa \\
      graph        &       &                                    &                                      &                             &                            &       &       &       &       &                            &         &                            &       &                            &       \\
      \midrule
      PP\_0.20\_0.10 &  0.15 &                               0.03 &                                 0.00 &                        0.09 &                       0.03 &  0.01 &  0.95 &  0.16 &  0.03 &                       0.00 &    0.54 &                       0.00 &  0.43 &  \cellcolor{green!50} 0.00 &  0.30 \\
      PP\_0.20\_0.15 &  0.15 &                               0.03 &                                 0.00 &                        0.14 &                       0.05 &  0.02 &  0.85 &  0.18 &  0.04 &                       0.00 &    0.57 &                       0.00 &  0.40 &  \cellcolor{green!50} 0.00 &  0.37 \\
      PP\_0.20\_0.20 &  0.14 &                               0.05 &                                 0.00 &                        0.18 &                       0.07 &  0.02 &  0.86 &  0.19 &  0.06 &  \cellcolor{green!50} 0.00 &    0.58 &                       0.00 &  0.42 &                       0.00 &  0.41 \\
      PP\_0.20\_0.25 &  0.15 &                               0.03 &                                 0.00 &                        0.24 &                       0.10 &  0.02 &  0.82 &  0.19 &  0.06 &                       0.00 &    0.48 &  \cellcolor{green!50} 0.00 &  0.44 &                       0.01 &  0.50 \\
\bottomrule
      \end{tabular}
    }
\end{table}

\end{document}